# Structural and Thermal Stability of Graphyne and Graphdiyne Nanoscroll Structures

*Daniel Solis[1,2], Daiane D. Borges[1], Cristiano F. Woellner[1], Douglas S. Galvão[1]*

[1]Applied Physics Department and Center for Computational Engineering & Sciences, University of Campinas - UNICAMP, Campinas-SP 13083-959, Brazil; [2]Univ. Grenoble Alpes, CEA, CNRS, Grenoble INP, INAC-SPINTEC, F-38000 Grenoble, France

ABSTRACT

Graphynes and graphdiynes are generic names for families of two-dimensional carbon allotropes, where acetylenic groups connect benzenoid-like hexagonal rings, with the co-existence of sp and $sp^2$ hybridized carbon atoms. The main differences between graphynes and graphdiynes are the number of acetylenic groups (one and two for graphynes and graphdiynes, respectively). Similarly to graphene nanoscrolls, graphyne and graphdiynes nanoscrolls are nanosized membranes rolled up into papyrus-like structures. In this work we investigated through fully atomistic reactive molecular dynamics simulations the structural and thermal (up to 1000K) stability of α,β,γ-graphyne and α,β,γ-graphdiyne scrolls. Our results show that stable nanoscrolls can be formed for all the structures investigated here, although they are less stable than corresponding graphene scrolls. This can be explained as a consequence of the higher graphyne/graphdiyne structural porosity in relation to graphene, which results in decreased π-π stacking interactions.

**INTRODUCTION**

With the materials science revolution created by the advent of graphene, there is a renewed interest in other 2D carbon allotropes, such as graphynes and graphdiynes[1-5]. These materials were predicted for the first time in 1987[1]. Graphynes and graphdiynes are generic names for families of two-dimensional carbon allotropes, where acetylenic groups connect benzenoid-like hexagonal rings (see Figures below), with the co-existence of sp and $sp^2$ hybridized carbon atoms. The main differences between graphynes and graphdiynes are the number of acetylenic groups (one and two for graphynes and graphdiynes, respectively). In principle, it is possible to create an almost infinite number of related structures, just increasing the number of acetylenic groups[1-5]. So far, most of the reported investigations has been focused on graphynes and graphdiynes[6,7]. Recent important advances have been possible due to the novel synthesis methodologies, which include metal-catalyzed cross-coupling reactions[8-10], template synthesis[11], and alkyne metasynthesis[12].

Nanoscrolls are structures obtained by rolling up nanosheeets into a papyrus-like topology[13] (Figure 1). In principle, any layered structure can be rolled up into scrolled ones, provided the necessary conditions for an energy assisted process and structural stability. In fact, many scrolls of different materials have already been obtained, such as graphene[14,15], $V_2O_5$[16], $H_2Ti_3O_7$[17] and 2D-hBN[18].

The scrolling process is a competition between elastic bending and van der Waals interactions. As the planar configurations are very stable, in order to curl it is necessary to provide external energy to induce the bending. The scroll formation can be a self-sustained process after a critical overlap area is reached, depending on the initial shape and the surface area of the nonosheet[19] (for a better visualization of the process, see video01 in the supplementary materials). The nanoscrolls are structures morphologically similar to multiwalled nanotubes, but with open ends[20]. Different techniques can be used to produce nanoscrolls. For example, graphite can be mixed with potassium oxide, separated with ethanol and then sonicated[14], which results in good quality carbon nanoscrolls. Another method involves the use of graphene sheets within a solution of isopropyl alcohol, which spontaneously scroll[20]. Recently, other experimental methods are being tried to order to produce scrolls of high quality and in large quantities[20]. The scrolled configuration can be even more stable that the planar configuration, as the increased on van der Waals interactions due to the scrolled process can be larger than the cost of the bending elastic energy to scroll it. The scroll topology for having ends open can be exploited in many applications, such as electroactuators (structures with mechanical charge dependent response), electronic or photoelectronic devices, as gas storage[19], etc. As graphyne and graphdiyne sheets have been already experimentally realized their scrolled configurations are feasible to obtain with our present day technologies.

In this work we have investigated the structural and thermal (up to 1000K) stability of graphyne and graphdiyne nanoscrolls through molecular dynamics (MD) simulations. We have considered α, β, and γ-type topologies (see Figure 2 and 5).

**METHODOLOGY**

Graphyne and graphdiyne nanoscrolls were investigated through MD simulations using the open source software named large-scale parallel molecular dynamics simulation code (LAMMPS)[21]. The atomistic pair interactions were described by the reactive force field ReaxFF [22]. Firstly, atomic coordinates of the planar sheets of graphyne (α, β, and γ-types) and graphdiyne (α, β, and γ-types) were built using the corresponding unit cell parameters. Subsequently, the sheets of width W and length L were rolled into an Archimedean spiral-shape (Figure 1b). The set of points that follows an Archimedean spiral are described by the following equation in polar coordinates:

$$r = a + b\theta, \qquad (1)$$

where $a$ is the starting point of the spiral or internal radius, $b$ is related to the separation among subsequent turns, *i.e.*, $2\pi b \sim 3.4$ Å is the interlayer distance, and $\theta$ is the number of turns.

Notice that since the sheet is finite, the value of *a* must vary accordingly to the number of turns. The suitable values of *a* were obtained through numerical calculations solving the arc length equation of the Archimedean spiral. A series of scrolls were generated with $\theta$ varying from 0 to $8\pi$. After generating the scroll coordinates, the sheets were rolled up using the Sculptor plugin implemented in the VMD software[23]. Due to the finite size of the sheets, to achieve a realistic and appropriate spiral shape, static calculations were initially performed to obtain the energy profiles as function of the number of turns and sheet dimensions (these calculations were performed at room temperature, T= 300 K).

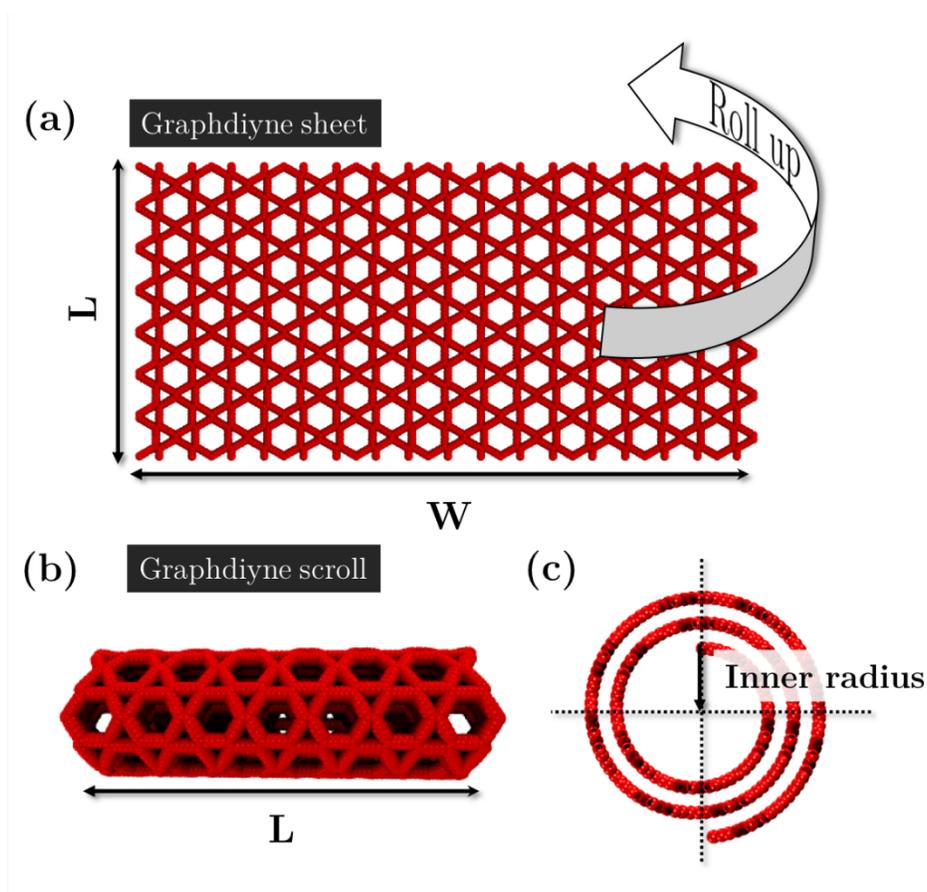

Figure 1: a) planar configuration of a sheet of width W and length L; b) formed nanoscroll of length L formed from rolled up the planar configuration displayed in (a); c) Top (cross-section view) of the nanoscroll displaying its inner radius, which is defined as the distance between the nanoscroll main axis and the closest line of atoms.

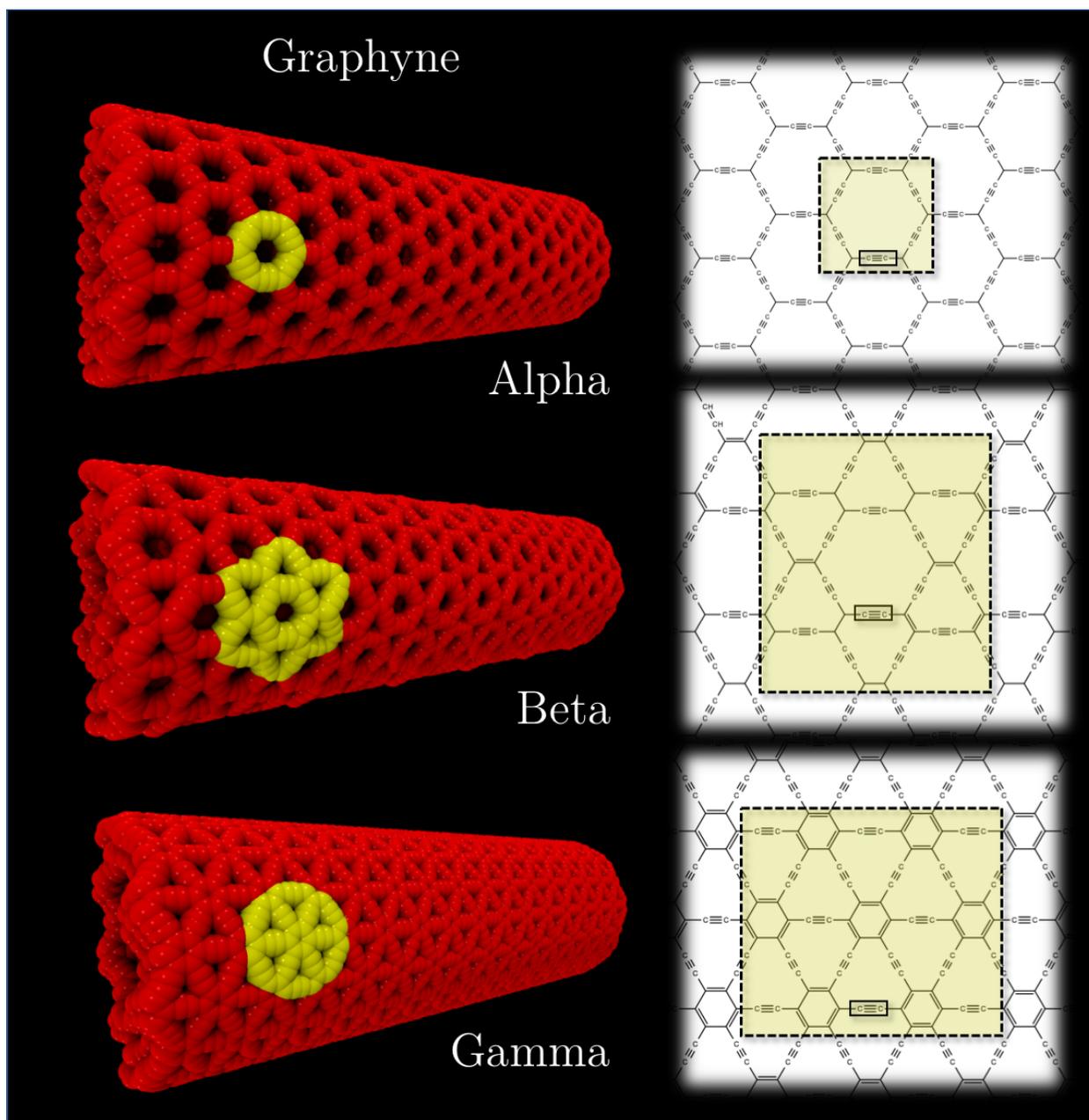

**Figure 2**: Schematics of the obtained α-, β- and γ-graphyne nanoscrolls (left side) from their corresponding planar configurations (right side). The highlighted regions (left and right) indicate the structural topological motifs.

For the dynamical analyses of the graphyne and graphdiyne nanoscrolls, we considered the parent planar structures sheets with width W~200 Å and length L~100 Å and two turns, this is θ

= 4π. These sheet dimensions are larger enough to allow many scrolled turns. A systematic study of the stability of nanoscrolls was carried out using a Nosé-Hoover thermostat and a canonical ensemble (NVT)[24,25]. For different temperatures in a range of 100 K up to 1000 K, the systems were thermalized until achieve equilibrium using a timestep of 0.2 fs. Due to the chosen initial number of turns and radius, all the nanoscrolls were out of the equilibrium. During the thermalization, these initial parameters changed. To properly quantify these variations, we used the internal nanoscroll radius values, which are defined as the distance between the line that pass through its center of mass (main axis) and the innermost line of atoms in the scroll (see Figure 1 c). It is important to point out that these values are obtained for each atom in the innermost line and averaged in order to take into account the thermal fluctuations.

**RESULTS AND DISCUSSION**

Graphyne:

Our MD results indicate that existence of stable scroll configurations for all different types and within the temperature range investigated here (up to 1000K). We did not observe broken bonds and/or the formation of new covalent bonds, even at high temperatures. Moreover, the scroll maintains its structure with inner radius values fluctuating around a constant value, as we can observe in the time evolution of inner radius displayed in Figure 3. The α-graphyne inner radius values fluctuates around ~5-6Å at temperatures below 500K (see Figure 3a). At high temperatures (>700K), these values increase to ~10-11Å. This effect of temperature is explained by the fact that the interlayer interaction energy that maintains the scroll cohesion/integrity is surpassed by the atomic kinetic energy, making the scroll 'breath'. The temperature has not the same effect in β- or γ-graphynes, where its inner radius is kept around 10-11 Å, independently of temperature. This radius is comparable to ~10 Å of graphene nanoscroll[13]. For β-graphyne in Figure 3b for the case of 100 K the stability radius value is around 8.5 Å and displays the smallest radius value among the different temperatures considered here. The same is similarly observed for α-graphyne. But differently to what occurs for α-graphyne, β-graphyne is less sensitive to temperature changes, showing already at 300 K a radius comparable to ~10 Å, the same value as at 500 K, 700 K, and 1000 K. In Figure 3c we can observe that the radius values present large amplitude oscillations, which disappears when the nanoscroll achieves its equilibrium configuration. The equilibrium radius values for all temperatures considered are very close and reflects the higher invariance with respect to temperature of the γ-graphyne in comparison to the other graphyne structures. This suggests that γ –graphyne is the most thermally stable configuration (see video02 in the supplementary materials).

In order to proper understand how the graphyne sheets change in terms of energy when they are rolled up, it is necessary to determine the balance between the energetic cost to bend the sheet, *i.e.*, the torsion energy and the energetic gain due the cohesion forces, *i.e.,* van der Waals (vdW) interactions, whose variation are the more pronounced. In Figure 4 we present the torsion energies (a) and the vdW energies (b) for each graphyne structure during its scrolling processes. Notice that the torsion energy increases while the scroll is being rolling up. The torsion energy cost is more relevant for the β-graphyne and less important for γ-graphyne, while α-graphyne represents an intermediate case. The vdW energy is zero until it reaches the value $2\pi$. For rounds infinitesimal larger than this value we start to have a sheet overlap and, consequently, a non-zero vdW interactions among the atoms of the overlapped surface can be observed (see Figure 4b for radius > $2\pi$). The energetic gain from vdW interactions is proportional to the overlap area, which means that increasing the scrolling implies increasing the vdW interaction energy gain.

In Figure 4c we present the sum of torsion and VDW per-atom energies. Starting the bending, but before the layer overlap occurs, the total energy of all the structures is increasing (because of the bending), clearly dominated by the torsion energy. When the graphyne is highly rolled up, the vdW gains surpasses the bending elastic cost and the scrolled configuration can be even more stable than the planar configuration. However, it should be stressed that in order to achieve the scrolled configuration this process needs to be energy assisted. Each graphyne-type possesses a different energy minimum and the most stable structure is the γ-graphyne. Even if β-graphyne is less porous (more vdW) than the α-graphyne, it is the least stable structure among the graphynes studied here. It should be stressed that the thermal stability ordering is not the same of structural stability one, since they have origin in different aspects of the scrolls.

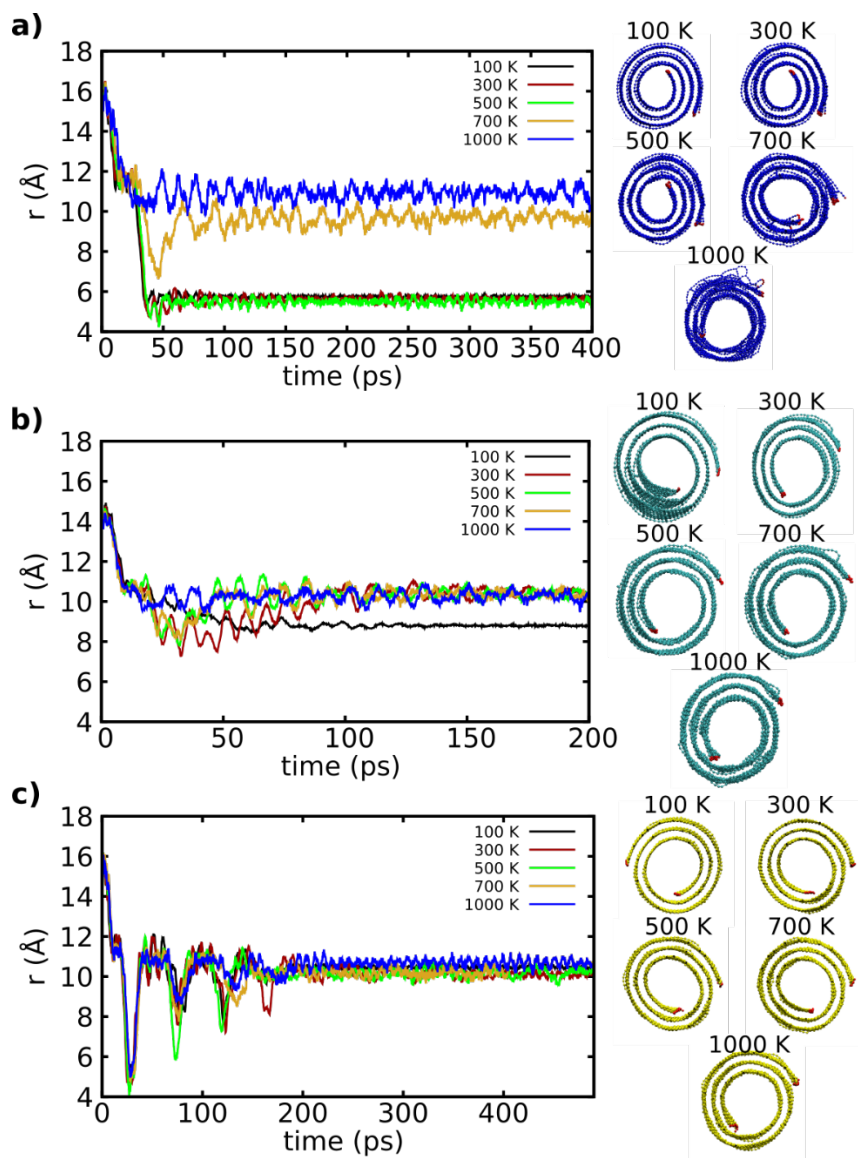

**Figure 3.** (Left) inner radius (r) values as a function of the simulation time for: a) α- graphyne; b) β-graphyne, and; c) γ-graphyne. (Right) MD snapshots of the obtained final configuration for temperatures 100 K, 300 K, 500 K, 700 K, and 1000 K, respectively. The red marks indicate the inner and outer edge of the corresponding nanoscroll.

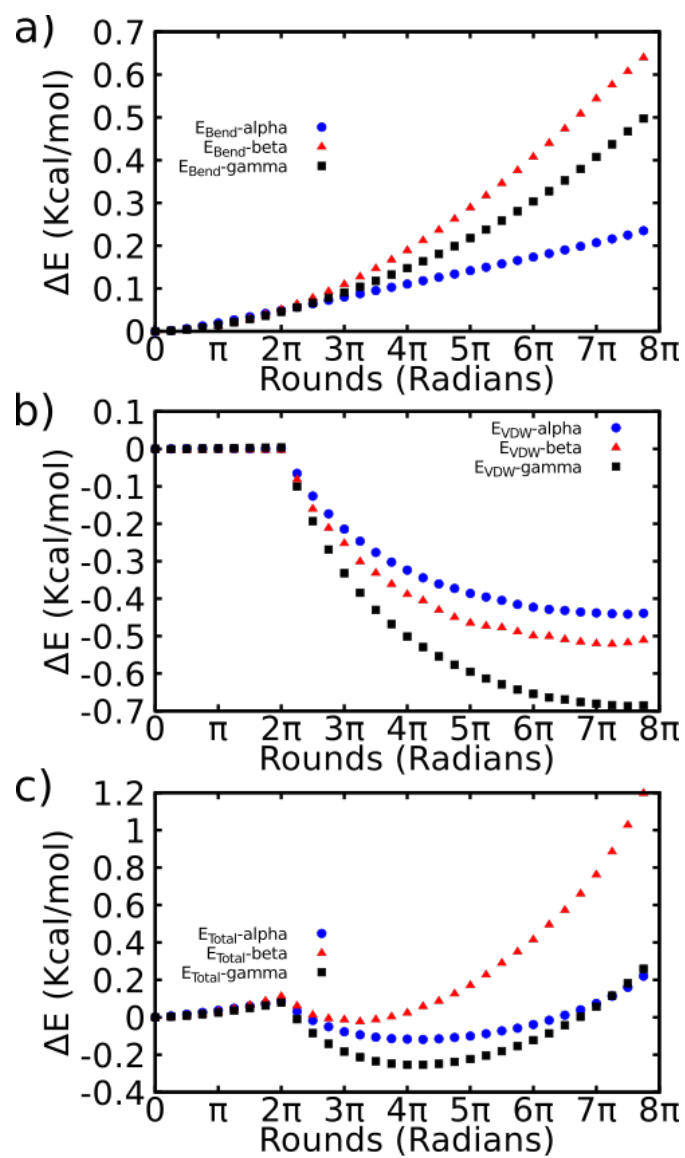

**Figure 4:** a) Torsion energy; b) van der Waals Energy, and; c) total energy for the static case for α-graphyne, β-graphyne and γ-graphyne.

## Graphdiynes

MD simulations were also carried out for three type of graphdiynes, considering the same structural parameters as for the graphynes, *i.e.,* width W~200 Å and height L~100 Å and initial configuration being rolled-up to 4π turns. The structures of the graphdiynes and their corresponding nanoscrolls are showed in Figure 5. The MD simulations have shown that these nanoscrolls also reached thermal stability for the temperature considered here (100K to 1000K). The total simulation time was between 400 ps and 600 ps. As mentioned above graphdiynes present twice as much acetylenic groups as compared to graphyne. Therefore, graphdiynes are more porous structures than graphynes. More porosity results in a decreased π-π stacking interactions, which results in a smaller number of turns when compared to graphynes (see video03 in the supplementary materials).

The graphdiyne scroll dynamics is similar to the graphynes ones, but the structural and thermal stability ordering are not the same. α-graphdiyne is the allotrope that exhibits the most evident elliptical-shaped structure, as shown in the snapshots of Figure 6a, as well as the graphdiyne structure with the largest radius value (a summary of the structural and average radius values for graphyne and graphdiyne scrolls are presented in Table 1).

Similarly, to the α-graphdiyne case, the β-graphdiyne (Figure 6b) nanoscroll also presents a cylindrical shape with an elliptic base. α-graphdiyne and β-graphdiyne are the most porous structures in this study with atomic densities of 0.238 atoms · Å$^{-2}$ and 0.259 atoms · Å$^{-2}$, respectively. Due to this, even at low temperatures, they form scrolls with a small number of turns (around 4π), and as a consequence these nanoscrolls have large radius values. Their larger deformation capability is directly related to the structural arrangements of their atoms, a feature that can be understood looking at the low torsion energies for these structures. In fact, their torsion energies are very close and the differences in the nanoscroll rolling comes mainly from the van der Waals interactions. The general trends are quite similar to the ones presented in Figure 4.

In Figure 6c we present the radius values variation and snapshots of the final configuration of γ-graphdiyne for different temperatures. Taking into account the γ-graphdiyne density and base shape low eccentricity, we can conclude that this structure is the least flexible and most stable allotrope among the graphdiynes. It possesses an average radius around 11 Å for all the temperatures considered here. This radius value is very similar to those found in the literature for graphene[13] and in the graphynes structures studied earlier.

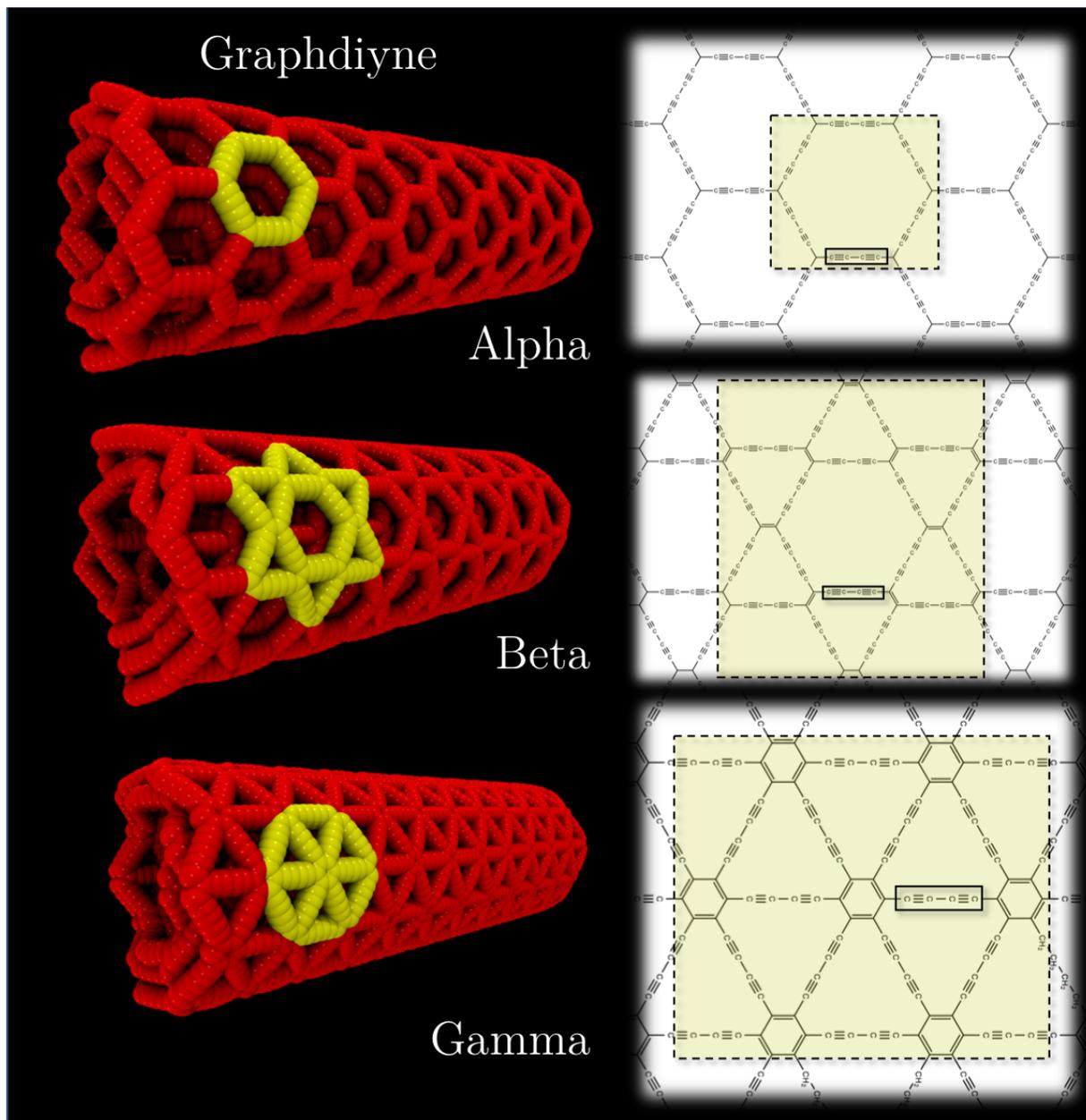

**Figure 5**: Schematics of the obtained α-, β- and γ-graphdiyne nanoscrolls (left side) from their corresponding planar configurations (right side). The highlighted regions (left and right) indicate the structural topological motifs.

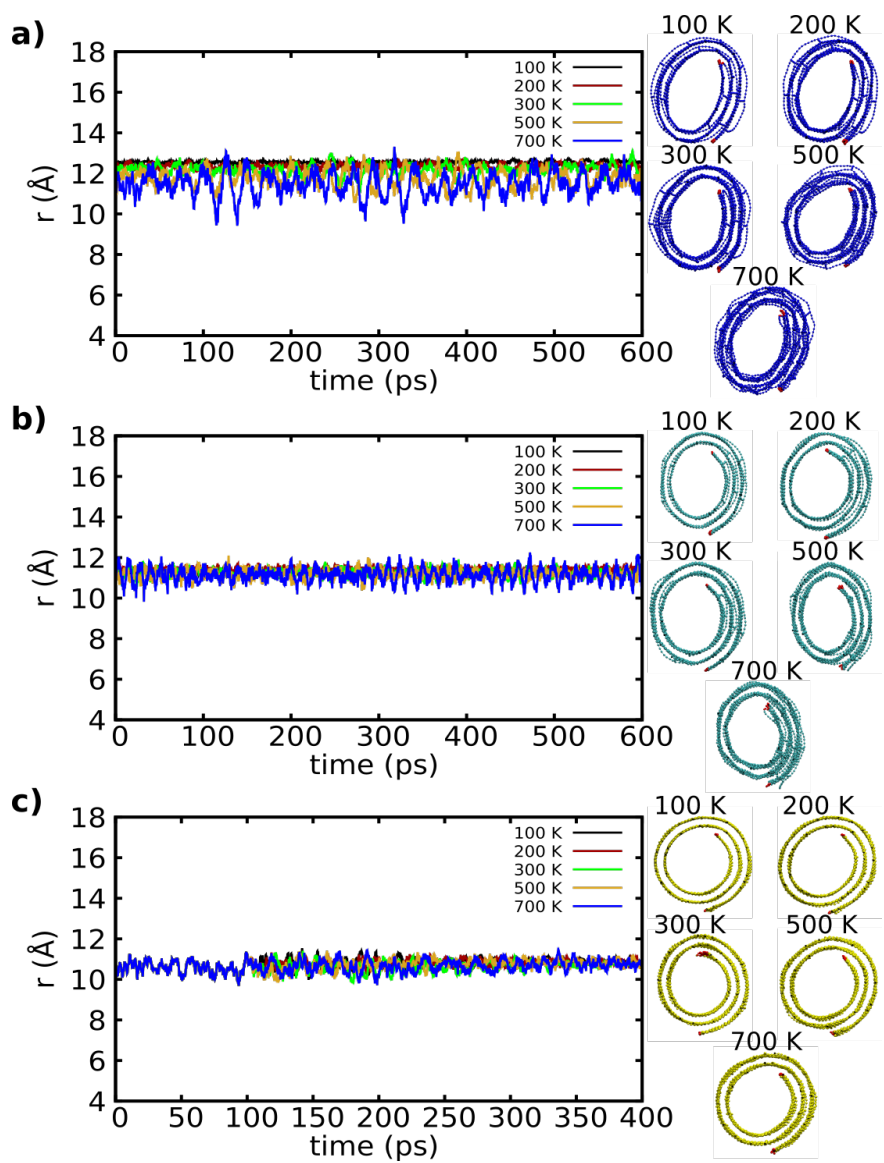

Figure 6: (Left) inner radius (r) values as a function of the simulation time for: a) α-graphdiyne; b) β-graphdiyne, and; c) γ-graphdiyne. (Right) MD snapshots of the obtained final configuration for temperatures 100 K, 200 K, 500 K, and 700 K, respectively. The red marks indicate the inner and outer edge of the corresponding nanoscroll.

**Table 1:** Atomic density and inner radius at temperatures 100 K, 300 K, 500 K, 700 K and 1000 K, for α-, β-, γ-, graphynes and α-, β-, γ-graphdiynes

| Structure | Atomic density atoms · Å$^{-2}$ | Radius (Å) | | | | | |
|---|---|---|---|---|---|---|---|
| | | 100 K | 200 K | 300 K | 500 K | 700 K | 1000 K |
| α-Graphyne | 0.238 | 5.77 | - | 5.62 | 5.32 | 9.57 | 11.33 |
| β-Graphyne | 0.259 | 8.82 | - | 10.33 | 10.32 | 10.09 | 10.14 |
| γ-Graphyne | 0.335 | 10.57 | - | 10.08 | 9.94 | 9.92 | 10.78 |
| α-Graphdiyne | 0.153 | 12.52 | 12.32 | 11.44 | 12.11 | 11.16 | - |
| β-Graphdiyne | 0.179 | 11.25 | 11.35 | 10.88 | 11.07 | 10.54 | - |
| γ-Graphdiyne | 0.253 | 10.86 | 10.72 | 10.86 | 10.76 | 10.99 | - |
| Graphene | 0.393 | - | - | - | - | - | - |

**CONCLUSIONS**

Graphynes and graphdiynes are generic names for families of two-dimensional carbon allotropes, where acetylenic groups connect benzenoid-like hexagonal rings. In this work we investigated through fully atomistic reactive (ReaxFF force field) molecular dynamics simulations the structural and thermal (up to 1000K) stability of α, β, γ graphyne and graphdiyne scrolls. These structures are graphynes and graphdiyne membranes rolled up into papyrus-like structures.

Our results show that stable (we did not observe broken bonds and/or the formation of new covalent bonds, even at high temperatures) nanoscrolls can be formed for all the structures investigated here, although there are less stable than corresponding graphene scrolls. This can be explained as a consequence of the higher graphyne/graphdiyne structural porosity in relation to graphene, which results in decreased π-π stacking interactions[26]. Interestingly, the structural and thermal stability ordering for these structures are not the same. Also, under heat excitation they undergo significant radial expansion, for instance the α-graphyne radius value can vary from 5.77 Å to 11.33 Å, which could be exploited in a large variety of applications, such as thermal actuators, sensors, etc. Considering that graphynes and graphdiyne membranes have been already experimentally realized, their scroll fabrication is feasible with our present day technology. We hope the present work can stimulate further works along these lines.

ASSOCIATED CONTENT

**Supporting Information**.

The following files are available free of charge: movies from MD simulations.

AUTHOR INFORMATION

**Corresponding Author**

Prof. Douglas S. Galvao

Applied Physics Department and Center for Computational Engineering & Sciences, University of Campinas - UNICAMP, Campinas-SP, 13083-959, Brazil

galvao@ifi.unicamp.br

**Author Contributions**

The manuscript was written through contributions of all authors. All authors have given approval to the final version of the manuscript.

ACKNOWLEDGMENT

This work was supported in part by the Brazilian Agencies CAPES, CNPq and FAPESP. The authors also thank the Center for Computational Engineering and Sciences at Unicamp for financial support through the FAPESP/CEPID Grant # 2013/08293-7.